\documentclass[]{article}
\input{psfig} %for CoRR
\newenvironment{pf}{\unskip{\bf Proof:}}{\unskip{\hfill $\Box$}}

\newcommand{\lemlab}[1]{\label{lemma:#1}}

\newcommand{\tablab}[1]{\label{tab:#1}}
\newcommand{\figlab}[1]{\label{fig:#1}}
\newcommand{\seclab}[1]{\label{section:#1}}

\newcommand{\lemref}[1]{\ref{lemma:#1}}

\newcommand{\tabref}[1]{\ref{tab:#1}}
\newcommand{\figref}[1]{\ref{fig:#1}}
\newcommand{\eqref}[1]{(\ref{eq:#1})}
\newcommand{\secref}[1]{\ref{section:#1}}

\newtheorem{theorem}{Theorem}
\newtheorem{lemma}{Lemma}
\newtheorem{conj}{Conjecture}
{\catcode`\@=11
\gdef\setft#1#2#3{%
\def\@oddfoot{
{\setbox0=\hbox{#1}
\setbox1=\hbox{#3}
\ifdim\wd0>\wd1
\dimen0=\wd0
\box0\hfil#2\hfil\hbox to\dimen0{\hfil\hfil\box1}
\else \dimen0=\wd1
\hbox to\dimen0{\box0\hfil }\hfil#2\hfil\box1 \fi
}}} }

\def\complaint#1{}
\def\withcomplaints{
\newcounter{mycomplaints}
\def\complaint##1{\refstepcounter{mycomplaints}%
\ifhmode%
\unskip%
{\dimen1=\baselineskip \divide\dimen1 by 2 %
\raise\dimen1\llap{\tiny -\themycomplaints-}}\fi%
\marginpar{\tiny [\themycomplaints]: ##1}}%
}

\usepackage{amssymb}
\usepackage{latexsym} %for CoRR

\title{\bf PushPush is NP-hard in 3D}
\author{%
Joseph~O'Rourke\thanks{
Dept. of Computer Science, Smith Col\-lege, North\-ampton, 
MA 01063, USA.
orourke@cs.\-smith\-.edu.
Research supported by NSF Grant CCR-9731804.
}
\\and 
The~Smith~Problem~Solving~Group\thanks{
Beenish Chaudry,
Sorina Chircu,
Elizabeth Churchill,
Sasha Fedorova,
Judy Franklin,
Biliana Kaneva,
Halley Miller,
Anton Okmianski,
Irena Pashchenko,
Ileana Streinu,
Geetika Tewari,
Dominique Thi{\'e}baut,
Elif Tosun.
}
}
\begin{document}
\maketitle
\begin{abstract}
We prove that a particular pushing-blocks puzzle is intractable in 3D.
The puzzle, inspired by the game {\em PushPush},
consists of unit square blocks on an integer lattice.
An agent may push blocks (but never pull them) in attempting
to move between given start and goal positions.
In the PushPush version, the agent can only push one block at
a time, and moreover, each block, when pushed, slides the
maximal extent of its free range.
We prove this version is NP-hard in 3D by reduction from SAT.
The corresponding problem in 2D remains open.
\end{abstract}

\section{Introduction}
There are a variety of ``sliding blocks'' puzzles whose
time complexity has been analyzed.
One class, typified by the 15-puzzle so heavily
studied in AI, permits an outside
agent to move the blocks.
Another class falls more under the guise of
motion planning.
Here a robot or internal agent plans a
path in the presence of movable obstacles.
This line was initiated by a paper of Wilfong~\cite{w-mppmo-91},
who proved NP-hardness of a particular version in which the
robot could pull as well as push the obstacles, which were
not restricted to be squares.
Subsequent work sharpened the class of problems by weakening
the robot to only push, never pull obstacles, and by 
restricting all obstacles to be unit squares.
Even this version is NP-hard~\cite{do-mpams-92}.

One theme in this research has been to establish stronger degrees
of intractability, in particular,   
to distinguish between NP-hardness and PSPACE-completeness, 
the latter being the stronger claim.  
The NP-hardness proved in~\cite{do-mpams-92} was 
strengthened to PSPACE\--completeness in a unfinished 
man\-u\-script \cite{bos-mpams-94}.  
More firm are the results on
Sokoban, a
computer game that restricts the pushing robot to only push one block at
a time, and requires the storing of (some or all) blocks into 
designated ``storage
locations.''
This game was proved NP-hard in~\cite{dz-sompp-95},
and PSPACE-complete by Culberson~\cite{c-spc-99}.  

Here we emphasize another theme: finding a nontrivial version of
the game that is {\em not\/} intractable.  To date only the most uninteresting
versions are known to be solvable in polynomial time, for example,
where the robot's path must be monotonic~\cite{do-mpams-92}.
We explore a different version,
again inspired
by a computer game, PushPush.  The key difference is that
when a block is pushed, it necessarily slides the full extent
of the available empty space in the direction in which it was
shoved. This further weakens the robot's control, and the
resulting puzzle has certain polynomial characteristics.
We prove it is intractable in 3D, but leave the question of whether
it is polynomial in 2D an open problem.

\section{Problem Classification}
\seclab{Classification}
The variety of pushing- block puzzles may be classified by
several characteristics:

\begin{enumerate}
\item Can the robot pull as well as push?
\item Are all blocks unit squares, or may they have different shapes?
\item Are all blocks movable, or are some fixed to the board?
\item Can the robot push more than one block at a time?
\item Is the goal for the robot to move from $s$ to $t$,
or is the goal for the robot to push blocks into storage locations?
\item Do blocks move the minimal amount, exactly how far they
are pushed, or do they slide the maximal amount of their
free range?
\item The dimension of the puzzle: 2D or 3D?
\end{enumerate}

If our goal is to find the weakest robot and most
unconstrained puzzle conditions that still lead to intractability, 
it is reasonable consider robots who can only push~(1),
and to restrict all blocks to be unit squares~(2), 
as in \cite{do-mpams-92,dz-sompp-95,c-spc-99}, for
permitting robots to pull, and permitting blocks of other shapes,
makes it relatively easy to construct intractable puzzles.
It also makes sense to explore the goal of simply finding a path~(5)
as in \cite{w-mppmo-91,do-mpams-92}, rather than
the more challenging task of
storing the blocks as in Sokoban~\cite{dz-sompp-95,c-spc-99}.

Restricting attention to these choices
still leaves a variety of possible problem definitions.
If the robot can only move one block at a time, then the
distinction between all blocks movable and some fixed disappears,
because 2x2 clusters of blocks are effectively fixed to a robot
who can only push one.
If all blocks are movable and the robot can push more than one
at a time, then the blocks should be confined to a
rectangular frame. 

The version explored in this paper superficially seems that it might
lend itself to a polynomial-time algorithm: the robot can only push
one block~(4), all blocks are pushable~(3), and finally, the
robot's control over the pushing is further weakened by condition~(6):
once pushed, a block slides (as without friction) the maximal
extent of its free range in that direction.
We show the problem is intractable in 3D, and discuss the 2D version
in the final section.

\section{Elementary Gadgets}
First we observe, as mentioned above, that any 2x2 cluster of 
movable blocks is forever frozen to a PushPush robot, for there
is no way to chip away at this unit.  This makes it easy to
construct ``corridors'' surrounded by fixed regions to guide
the robot's activities.  We will only use corridors of
width~1 unit, with orthogonal junctions of degree two, three,
or four.  We can then view a particular PushPush puzzle as
an orthogonal graph, whose edges represent the corridors, understood
to be surrounded by sufficiently many 2x2 clusters to render any
movement outside the graph impossible.
We will represent movable blocks in the corridors or at corridor
junctions as circles.

We start with three elementary gadgets.

\subsection{One-Way Gadget}

A ``one-way'' gadget is shown in
Fig.~\figref{One-Way}a.
It has these obvious properties:
%%%%%%%%%%%%%%%%Figure Begin
\begin{figure}[htbp]
\begin{center}
\ \psfig{figure=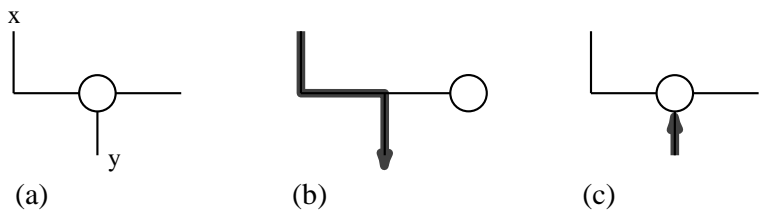,width=10cm}
%\fbox{figure=One-Way.eps}
\end{center}
\caption{One-Way gadget: permits passage from $x$ to $y$ but
not from $y$ to $x$.}
\figlab{One-Way}
\end{figure}
%%%%%%%%%%%%%%%%Figure End

\begin{lemma}
In a One-Way gadget,
the robot may travel from point $x$ to point $y$,
but not from $y$ to $x$.
(After travelling from $x$ to $y$, however,
the robot may subsequently return from $y$ to $x$.)
\lemlab{One-Way}
\end{lemma}
\begin{pf}
The block at the degree-three junction may be pushed into
the storage corridor when approaching from $x$,
as illustrated in Fig.~\figref{One-Way}b,
but the block may not be budged when approaching from $y$
(Fig.~\figref{One-Way}c).
\end{pf}

\subsection{Fork Gadget}
The fork gadget shown in
Fig.~\figref{Fork}a presents the robot with a binary choice,
the proverbial fork in the road:
%%%%%%%%%%%%%%%%Figure Begin
\begin{figure}[htbp]
\begin{center}
\ \psfig{figure=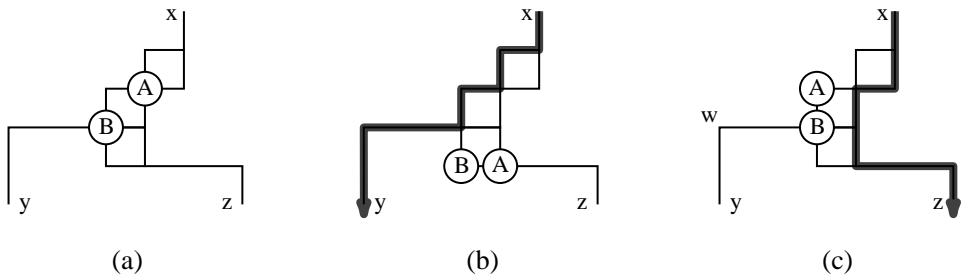,width=5in}
%\fbox{figure=Fork.eps}
\end{center}
\caption{Fork gadget: Robot may pass (b) from $x$ to $y$
or (c) from $x$ to $z$, but each seals off the other possibility.}
\figlab{Fork}
\end{figure}
%%%%%%%%%%%%%%%%Figure End

\begin{lemma}
In a Fork gadget,
the robot may travel from point $x$ to $y$, or from
$x$ to $z$, but if it chooses the former it cannot later
move from $y$ to $z$, and if it chooses the latter it cannot
later move from $z$ to $y$.
(In either case, the robot may reverse its original path.)
\lemlab{Fork}
\end{lemma}
\begin{pf}
Fig.~\figref{Fork}b shows the only way for the robot to pass
from $x$ to $y$.  Now the corridor to $z$ is permanently sealed
off.
Fig.~\figref{Fork}c shows the only way to move from $x$ to $z$.
Here any attempt later to access the corridor leading to $y$
will necessarily push block $B$ to corner $w$, sealing off $y$.
\end{pf}

Note that in both these gadgets, the robot may reverse its
path, a point to which we will return in Section~\secref{2D}.

\subsection{3D Crossover Gadget}
Crossovers are trivial in 3D, as shown in 
Fig.~\figref{3D.crossover}.
%%%%%%%%%%%%%%%%Figure Begin
\begin{figure}[htbp]
\begin{center}
\ \psfig{figure=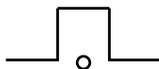}
%\fbox{figure=3D.crossover.eps}
\end{center}
\caption{3D crossover. The central small circle is a wire
orthogonal to the plane of the figure.}
\figlab{3D.crossover}
\end{figure}
%%%%%%%%%%%%%%%%Figure End

\section{Variable-Setting Component}
The robot first travels through a series of variable-setting
components, each of which follows the structure shown in
Fig.~\figref{Variable}: a Fork gadget, followed by two paths,
labeled {\sc t} and {\sc f}, each
with attached {\em wires} exiting to the right,
followed by a re-merging of the
the {\sc t} and {\sc f} paths via One-Way gadgets.
3D crossovers are illustrated in this and subsequent figures
by broken-wire underpasses.
%%%%%%%%%%%%%%%%Figure Begin
\begin{figure}[htbp]
\begin{center}
\ \psfig{figure=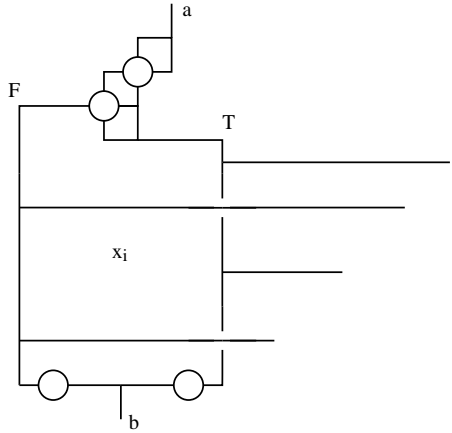,height=6cm}
%\fbox{figure=Variable.eps}
\end{center}
\caption{(a) Variable $x_i$ component.}
\figlab{Variable}
\end{figure}
%%%%%%%%%%%%%%%%Figure End
\begin{lemma}
The robot may travel from $a$ to $b$ only by choosing either the
{\sc t}-path, or the {\sc f}-path, but not both.
Whichever {\sc t/f}-path is chosen allows the robot to travel down any wires
attached to that path, but down none of the wires attached to the other
path.
\lemlab{Variable}
\end{lemma}
\begin{pf}
The claims follow directly from
Lemma~\lemref{Fork}
and
Lemma~\lemref{One-Way}.
\end{pf}

\section{Clause Component}
The clause component 
shown in Fig.~\figref{Clause}a
cannot be traversed unless one or more blocks are pushed in
from the left along the attached horizontal wires.
%%%%%%%%%%%%%%%%Figure Begin
\begin{figure}[htbp]
\begin{center}
\ \psfig{figure=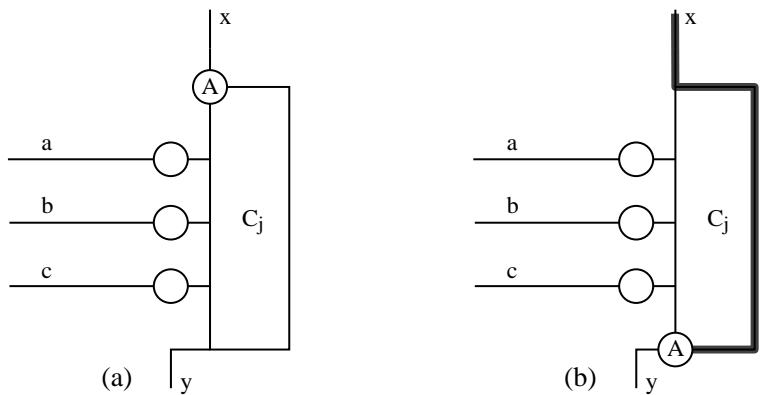,width=10cm}
%\fbox{figure=Clause.eps}
\end{center}
\caption{(a) Clause $C_j$ component.}
\figlab{Clause}
\end{figure}
%%%%%%%%%%%%%%%%Figure End
\begin{lemma}
The robot may only pass from $x$ to $y$ of a clause component
if at least one block is pushed into it along an attached wire
($a$, $b$, or $c$ in Fig.~\figref{Clause}a).
\lemlab{Clause}
\end{lemma}
\begin{pf}
Block $A$ is necessarily pushed by the robot starting at $x$.
This block will clog exit at $y$ (Fig.~\figref{Clause}b)
unless its sliding is stopped
by a block pushed in on an attached wire.
\end{pf}

\section{Complete SAT Reduction}
The complete construction for four clauses
$C_1 \wedge C_2  \wedge C_3 \wedge C_4$
is shown in 
Fig.~\figref{3D.SAT}.
Two versions of the clauses are shown in the figure:
an unsatisfiable formula (the dark lines),
and a satisfiable formula (including the shaded $x_2$ wire):

\begin{eqnarray}
&(x_1 \vee x_2) \wedge  (x_1 \vee \sim x_2) \wedge  (\sim x_1 \vee x_3) \wedge (\sim x_1 \vee \sim x_3) \\
&(x_1 \vee x_2) \wedge  (x_1 \vee \sim x_2) \wedge  (\sim x_1 \vee x_2 \vee x_3) \wedge (\sim x_1 \vee \sim x_3)
\end{eqnarray}
Here we are using $\sim x$ to represent the negation of the variable $x$.

%%%%%%%%%%%%%%%%Figure Begin
\begin{figure}[htbp]
\begin{center}
\ \psfig{figure=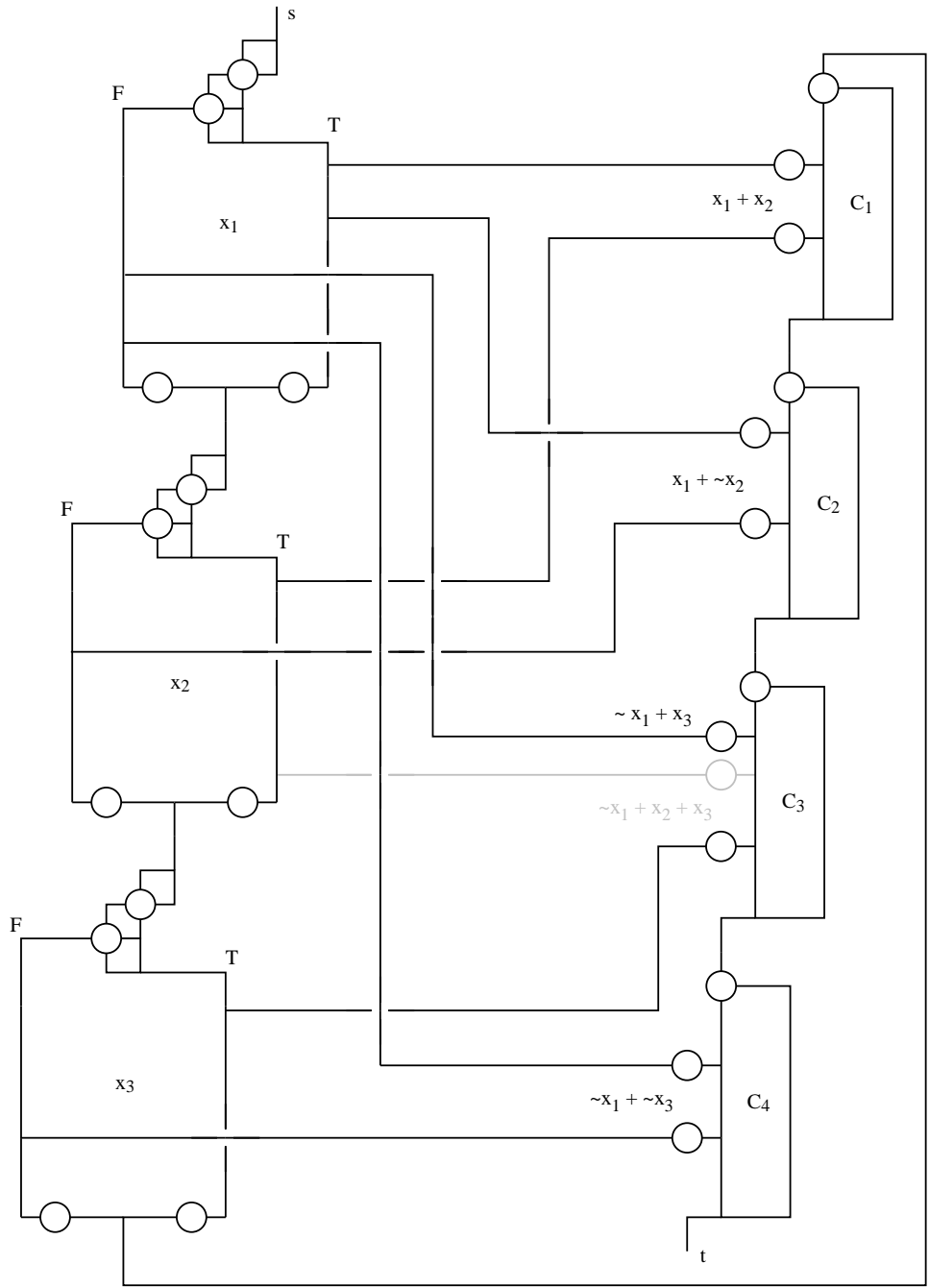,width=5in}
%\fbox{figure=3D.SAT.eps}
\end{center}
\caption{Complete construction for the formulas in Eq.~(1) and Eq.~(2)
(including the shaded portion).}
\figlab{3D.SAT}
\end{figure}
%%%%%%%%%%%%%%%%Figure End
A path from $s$ to $t$ in the satisfiable version is illustrated in
Fig.~\figref{3D.SAT.path}.
%%%%%%%%%%%%%%%%Figure Begin
\begin{figure}[htbp]
\begin{center}
\ \psfig{figure=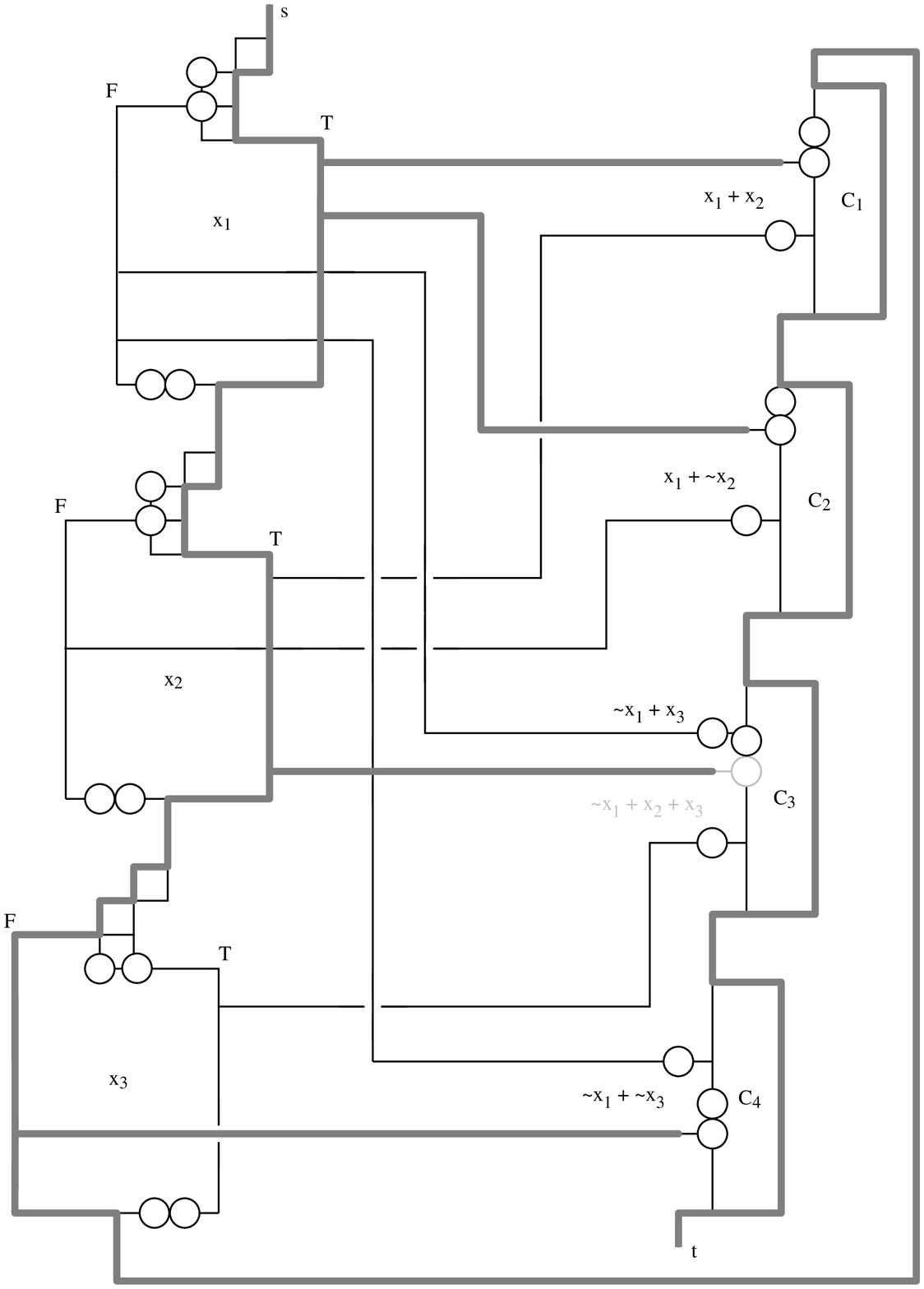,width=5in}
%\fbox{figure=3D.SAT.path.eps}
\end{center}
\caption{Solution path for Eq.~(2).}
\figlab{3D.SAT.path}
\end{figure}
%%%%%%%%%%%%%%%%Figure End

\begin{theorem}
PushPush is NP-hard in 3D.
\end{theorem}
\begin{pf}
The construction clearly ensures,
via Lemmas~\lemref{Variable} and~\lemref{Clause},
that if the simulated Boolean
expression is satisfiable, there is a path from $s$ to $t$,
as illustrated in Fig.~\figref{3D.SAT.path}.
For the other direction, suppose the expression is unsatisfiable.
Then the robot can reach $t$ only by somehow ``shortcutting'' the
design.  The design of the variable components ensures that
only one of the {\sc t/f} paths may be accessed.
The crossovers ensure there is no ``leakage'' between wires.
The only possible thwarting of the design would occur if the
robot could travel from a clause component back to set a variable
to the opposite Boolean value.  But each variable-clause
wire contains a block that prevents any such leakage.
\end{pf}

\section{PushPush in 2D}
\seclab{2D}

It is an intriguing question whether the 2D version of this
problem is intractable.  One result in this direction is easy to obtain:

\begin{theorem}
The storage version of PushPush is NP-hard in 2D.
\end{theorem}
\begin{pf}
In the storage version of PushPush,\footnote{
        This, incidentally, is the actual design of the computer game.
}
the robot must fill certain storage locations with blocks,
as in Sokoban.  It is then easy to obtain an NP-hardness proof
along the lines of the NP-hardness proof of Dor and Zwick~\cite{dz-sompp-95}.
Rather than reducing from SAT, reduce from ``Planar 3-SAT''~\cite{l-pftu-82}.
This, together with the storage requirement, removes the need for
any crossovers.  The construction can then follow the design
as in Fig.~\figref{3D.SAT}.  Details 
are similar to those in~\cite{dz-sompp-95} and will not be presented.
\end{pf}

\noindent
The reason that Planar 3-SAT does not help for the path version of
PushPush is that crossovers are still needed to thread the
clause components together into a single path.
And it seems that the PushPush conditions are too weak to 
construct the required crossover gadget:

\begin{conj}
No general crossover gadget can be constructed in 2D PushPush.
\end{conj}
Such a gadget would permit two wires to cross, but would prevent
leakage from one to the other, just as if it were a 3D crossover.
One reason this seems impossible is this:

\begin{conj}
No permanent one-way gadget can be constructed in 2D PushPush.
\end{conj}
Note that the the properties of the One-Way gadget in Fig.~\figref{One-Way}
are destroyed by passage of the robot, after which it
becomes a two-way street.

We conclude by summarizing 
in Table~\tabref{results}
previous work according to
the classification scheme offered in Section~\secref{Classification}.
The first four lines show previous results.
The next two are the results from this paper.
And the last two lines pose two open problems,
one raised here, the other in~\cite{do-mpams-92}:
Is PushPush (path version) intractable in 2D?
And is the problem where all blocks are movable and
the robot can push $k$ blocks, sliding the minimal amount,
intractable in 2D?

\begin{table}[htbp]
\begin{center}
\begin{tabular}{| c | c | c | c | c | c | c | c |}
        \hline
1 & 2 & 3 & 4 & 5 & 6 & 7 & \mbox{} \\
{\em Push?}
        & {\em Blocks}
        & {\em Fixed?}
        & {\em \#}
        & {\em Path?}
        & {\em Sliding}
        & {\em Dim}
        & {\em Complexity}
        \\ \hline \hline
pull
        & L
        & fixed
        & $k$
        & path
        & min
        & 2D
        & NP-hard \cite{w-mppmo-91}
        \\ \hline
push
        & unit
        & fixed
        & $k$
        & path
        & min
        & 2D
        & NP-hard \cite{do-mpams-92}
        \\ \hline
push
        & unit
        & movable
        & $1$
        & storage
        & min
        & 2D
        & NP-hard \cite{dz-sompp-95}
        \\ \hline
push
        & unit
        & movable
        & $1$
        & storage
        & min
        & 2D
        & PSPACE \cite{c-spc-99}
        \\ \hline \hline
push
        & unit
        & movable
        & $1$
        & path
        & {\em max}
        & {\em 3D}
        & NP-hard
        \\ \hline
push
        & unit
        & movable
        & $1$
        & storage
        & {\em max}
        & 2D
        & NP-hard
        \\ \hline \hline

push
        & unit
        & movable
        & $1$
        & path
        & {\em max}
        & 2D
        & open
        \\ \hline
push
        & unit
        & movable
        & $k$
        & path
        & min
        & 2D
        & open \cite{do-mpams-92}
        \\ \hline
\end{tabular}
\end{center}
\caption{Pushing block problems.}
\tablab{results}
\end{table}

\paragraph{Acknowledgements.}
We thank Erik Demaine, Marty Demaine, and Therese Biedl for 
helpful discussions.

\bibliographystyle{alpha}
\bibliography{/home1/orourke/bib/geom/geom}
\end{document}